\documentclass[twocolumn,aps,prl,amssymb,superscriptaddress]{revtex4-1}

\usepackage{graphicx}

\def\tPA{\emph{t}-PA} 
\def\nn{$(n,n)$}

\begin{document}

\title{Peierls Instability in Carbon Nanotubes}
\author{Guillaume Dumont}
\affiliation{D\'epartement de physique et Regroupement qu\'eb\'ecois sur les mat\'eriaux de pointe (RQMP),}
\author{Paul Boulanger}
\affiliation{D\'epartement de physique et Regroupement qu\'eb\'ecois sur les mat\'eriaux de pointe (RQMP),}
\author{Michel C\^ot\'e}
\email{Michel.Cote@umontreal.ca}
\affiliation{D\'epartement de physique et Regroupement qu\'eb\'ecois sur les mat\'eriaux de pointe (RQMP),}
\author{Matthias Ernzerhof}
\affiliation{D\'epartement de chimie,
Universit\'e de Montr\'eal, C. P. 6128 Succursale Centre-ville,
Montr\'eal (Qu\'ebec) H3C 3J7, Canada}

\begin{abstract}
We present a first-principles study of Peierls distortions in
\emph{trans}-polyacetylene, polyacene, and armchair $(n,n)$ carbon
nanotubes. Our findings suggest that the ground-state geometries of armchair
$(n,n)$ carbon nanotubes, with $n$ up to 6, exhibit a Peierls distortion as
it is found for \emph{trans}-polyactetylene. In contrast to previous studies
in which no Peierls distortion is found with conventional local and
semi-local density functionals, we use a hybrid functional whose
exact-exchange admixture has been specifically optimized for the problem at
hand.
\end{abstract}

\maketitle

The debate  whether carbon nanotubes should exhibit a Peierls instability arose shortly after their discovery \cite{Mintmire:1992}. \emph{Trans}-polyacetylene (\emph{t}-PA) is the classic example of Peierls instability and since $(n,n)$ armchair nanotubes can be considered as a parallel arrangement of polyacetylene chains along the tube axis, it is natural to suspect that such a deformation is also possible in these systems. Studies \cite{Mintmire:1992,Tanaka:1996,Sedeki:2000} of the electron-phonon coupling using model Hamiltonians predict a Peierls transition with low critical temperature. Whereas earlier investigations based on density-functional theory (DFT) found a symmetric geometry, more recent investigations using denser grids or refined zone-folding techniques \cite{Bohnen:2004,Connetable:2005,Piscanec:2007} also reveal a Peierls instability for small nanotubes. All these studies found an instability at $q=2k_F$ which results in a reduction of the translational symmetry. The corresponding distortion is quite different than the deformation in \emph{t}-PA because at least three unit cells are required to describe it. According to Ref. \cite{Connetable:2005}, many more unit cell could be necessary, because they find $k_F$ to be incommensurate with the unit cell. The phonon spectrum obtained with DFT calculations \cite{Bohnen:2004,Connetable:2005,Piscanec:2007} also indicates a softening at $q=0$ which corresponds to a deformation that only doubles the unit cell, i.e. a deformation that is quite similar to the one found in \emph{t}-PA.

However, it has been shown \cite{tPA_white,tPA_boehm} that the standard local density (LDA) and generalized gradient approximations (GGAs) are unable to reproduce the experimental geometry of trans-polyacetylene \cite{tPA_exp_1,tPA_exp_2,tPA_exp_3,tPA_exp_4}, underestimating the bond length alternation (BLA) by 70-80\%. Thus, in order to reproduce the correct ground state of \emph{t}-PA from a first-principles method, theories which treat exchange and correlation more accurately have to be employed \cite{tPA_suhai,tPA_karpfen,tPA_kertesz}. Hybrid functionals, in which a part of the approximate DFT exchange is replaced by exact exchange, have proven to give a more accurate estimat of the BLA in \emph{trans}-polyacetylene \cite{tPA_suhai}.

The previous DFT studies mentioned above used local and semi-local density  functionals but in light of results for \emph{t}-PA, it seems important to consider the effects of hybrid functionals on armchair carbon nanotubes. Notably, it could change the relative strength of the instability at $q=0$ compared to $q=2k_F$. Understanding the nature of the instability is crucial since it could well be related to the superconductivity found in carbon nanotubes \cite{Tang:2001,Takesue:2006}.

In this article, we report an \emph{ab initio} study of Peierls instabilities in \emph{trans}-polyacetylene, polyacene (PAc), and $(n,n)$ armchair carbon nanotubes and highlight the importance of including exact exchange in the functional to obtain a dimerized structure.


{\em Method:} All calculations were done with the Kohn-Sham approach of DFT using gaussian basis sets, as implemented in the periodic boundary conditions (PBC) module of the \textsc{Gaussian}03 package \cite{g03}. We employed the 6-311G and 6-31G basis sets for the polymers and the nanotubes respectively, and the Brillouin zones were sampled with over 200 $\mathbf{k}$-points. Atomic positions and unit cell lengths of the tubes and polymers have been optimized. Several exchange-correlation functionals have been put to work in this study, namely the PBE \cite{pbepbe}, B3LYP \cite{b3lyp,b3lyp2}, HSE \cite{hse}, as well as the PBE hybrid \cite{pbe1pbe}, referred thereafter as hPBE, and a generalization of the latter:
\begin{equation}\label{eq:PBEXC}
E_{XC} = (1-x) \Delta E^{\textrm{PBE}}_X+xE^{\textrm{exact}}_X + E^{\textrm{PBE}}_C,
\end{equation}
where $x$ is the fraction of exact exchange included in the functional. The PBE and hPBE functionals correspond to $x=0$ and $x=0.25$, respectively. We will refer, hereafter, to the various flavors of this functional by specifying only the fraction of exact exchange $x$. We would like to stress that the optimal amount of exact exchange mixing depends on the system of interest \cite{Ernzerhof1996499,Ernzerhof1998}. Therefore, the "best" hybrid functional for the problem at hand contains one unknown parameter. Furthermore we should mention that the implementation of the hybrid methods in the GAUSSIAN does not strictly adhere to the Kohn-Sham approach since the non-local Hartree-Fock exchange operator is employed instead of the local exchange potential. Experience indicates that this modification has essentially no effect on the predicted ground-state properties \cite{teale:034101}.   
%
The HSE functional used in this study includes 25\% of $E^{\textrm{exact}}_X$ and the Coulomb interaction is screened at a distance of 3.5 \AA (see the reference by Heyd {et al.} \cite{hse} for details of the implementation).

\begin{figure}
    \centering
    \includegraphics[width=0.45\columnwidth]{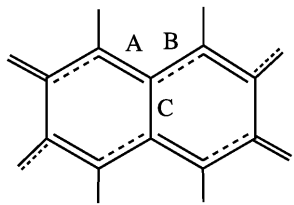}
	\caption{\label{fig:type_of_dist}Most stable configurations obtained for PAc and $(n,n)$ CNTs with $x>0$. Here the double, aromatic and single bonds notation is used to emphasize that the bond lengths are ordered as follows $A<B<C$.}
\end{figure}

{\em Distortion \emph{vs} exact exchange:} In this section we report the bond length alternation along the system axis, defined as the difference between the long $B$ and short $A$ bonds, and the associated band gap obtained using different exchange-correlation functionals. The systems considered are  \emph{trans}-polyacetylene, polyacene, and carbon nanotubes. The distortion found to be the most stable in PAc and $(n,n)$ tubes is depicted in Fig.~\ref{fig:type_of_dist}. These systems are very similar to that of a parallel arrangement of distorted \emph{t}-PA chains along the tube (or polymer) axis. The BLA of neighboring chains exhibits a phase difference corresponding to half a cell. Hence, the distorted geometry is found to be a Kekul\'e-type structure, the bond lengths being all different and are ordered as follows: $A<B<C$ (see Table \ref{tab:tubes_geom_etot}). Note that it is also different from what was reported in \cite{Connetable:2005}, but it is very similar to that of \cite{Mintmire:1992} and the one found in \cite{Figge:2001}, which also found a Kekul\'e-type structure.

Table \ref{tab:XCfuncs} shows the BLA (denoted by $\Delta a$) and the band gap, $E_g$, obtained with the B3LYP (20\% of $E^{\textrm{exact}}_X$), hPBE (25\% of $E^{\textrm{exact}}_X$) and HSE (25\% of screened $E^{\textrm{exact}}_X$) functionals for \tPA, PAc and the (3,3) nanotube. As expected, the band gap is proportional to the amount of exact exchange included in the functional and so is the BLA. This stems from the fact that Hartree-Fock theory tends to overestimate band gaps and over localize electrons. Hence, an appropriate choice of the parameter $ x $  can recover the correct BLA. We see from Table \ref{tab:XCfuncs} that although the HSE functional yields a  dimerized structure for \emph{t}-PA and PAc, the magnitude of the BLA is less than that obtained with functionals which include the full extent of the exchange interaction suggesting that the medium to long-range exchange interaction increases the dimerization in these systems. In particular, for the $(3,3)$  nanotube, HSE predicts a symmetric geometry with a very small gap. The importance of this long-range interaction could explain the failure of purely local- and semi-local functionals, such as LDA and PBE.
\begin{table}
	\begin{center}
		\begin{tabular}{|l|cc|cc|cc|}
\cline{2-7}
\multicolumn{1}{c|}{}           & \multicolumn{ 2}{c}{B3LYP} & \multicolumn{ 2}{c}{hPBE} & \multicolumn{ 2}{c|}{HSE} \\
\cline{2-7}
\multicolumn{1}{c|}{}        & $\Delta a$ &   $E_g$  & $\Delta a$ &   $E_g$  & $\Delta a$  &   $E_g$  \\
\multicolumn{1}{c|}{}        & ($10^{-3}$ \AA) &  (eV) & ($10^{-3}$ \AA) &  (eV) & ($10^{-3}$ \AA) & (eV) \\
\hline
		        \emph{t}-PA &   56.4 &       1.23 &   59.0 &       1.49 &   49.6 &       0.84 \\

		       PAc &   24.8 &       0.44 &   26.7 &       0.55 &   17.0 &       0.17 \\

		     (3,3) &   4.13 &       0.25 &   4.97 &       0.36 &   0.0 &       0.05 \\
\hline		
		\end{tabular}
	\end{center}
	\caption{Calculated distortions and band gaps for \tPA, PAc, and a (3,3) tube with different exchange-correlation functionals.}
	\label{tab:XCfuncs}
\end{table}

\begin{figure}
	\begin{center}
		\includegraphics[width=0.95\columnwidth]{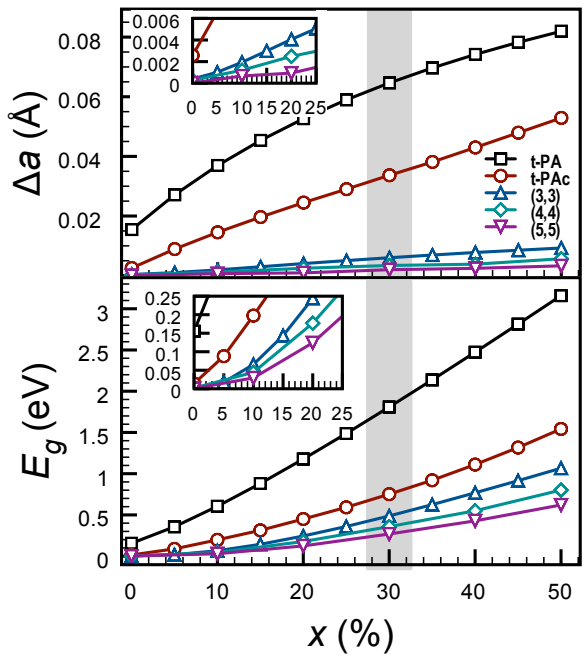}
	\end{center}
	\caption{(color online) Dimerization amplitude (top panel) and band gap (bottom panel) as a function of the percentage of exact exchange ($x$) included in the functional. The shaded area indicates the range of $x$ yielding the best agreement between calculated and experimental \emph{t}-PA bond lengths. The insets zoom in on the small-$x$ region.}
	\label{fig:dimer_amp}
\end{figure}

To systematically study the exact exchange dependence of the dimerization amplitude and band gap, we used the functional of Eq.~\ref{eq:PBEXC} and varied $x$ between 0 and 0.5. Fig.~\ref{fig:dimer_amp} shows the dimerization amplitude (top panel) and the band gap (bottom panel) of \tPA, PAc, and \nn~CNTs, for $n$ ranging from 3 to 5, as a function of $x$. The geometries were optimized for each value of $x$. The reported band gaps are differences in Kohn-Sham eigenvalues and should be treated only as indications of a trend since the electron-hole interaction is not included and it has been shown to be important in nanotubes \cite{Spataru:2004,Wang:2007}. The BLA in the tubes scales linearly with the amount of exact exchange, while the band gap follows a power law $x^m$ with $m > 1$. This figure also shows that the density only PBE ($x=0$) functional is unable to reproduce the experimental BLA of 0.08 \AA~nor the band gap of 1.5 eV \cite{tPA_exp_5,tPA_exp_6} of \emph{trans}-polyacetylene and predicts all $(n,n)$ CNTs to be conductors. The increase in the band gap as a function $x$ is caused by the addition of exact exchange included in the functional and by the geometry of the tube that is more dimerized as $x$ increased.

{\em Optimized geometry of the tubes:} To study the Peierls mechanism in more detail, we use the amount of exact exchange that best describes the \emph{trans}-polyacetylene ground state geometry since the tubes and PAc can be viewed as an arrangement of \tPA\ chains. For PAc the best agreement with the experimental geometry was obtained for $x=0.3$. We prefer this hybrid functional over the B3LYP (or a generalization of it) because it recovers the homogenous electron gas limit. In Table \ref{tab:tubes_geom_etot}, the optimized bond lengths, differences in total energy with respect to the symmetrical case and the band gaps are reported for \tPA, PAc and $(n,n)$ obtained with $x=0.3$. From these results it is clear that the distortion becomes less significant as the tube diameter increases. Also, all bond lengths seem to converge to the same value of 1.421 \AA\, which corresponds to the hPBE$(x=0.3)$ value for  the C--C bond length in graphite.

\begin{table}
	\begin{center}
				\begin{tabular}{|c|r|r|r|cc|c|c|}
					\cline{2-4}			
				 	\multicolumn{1}{c}{}           & \multicolumn{ 3}{|c|}{Bond lengths (\AA)}  \\
				\hline
				      $n$ & \multicolumn{1}{|c|}{A} & \multicolumn{1}{|c|}{B} & \multicolumn{1}{|c|}{C} & \multicolumn{ 2}{c|}{$\Delta E_\textrm{tot}$ (meV, K)} &  $E^{A<B}_g$ (eV)\\
				\hline
				        \emph{t}-PA &      1.361 &      1.426 &    \multicolumn{1}{c|}{---} & 27.11 &  314.56 &        1.66     \\

				       PAc &      1.383 &      1.416 &      1.460 &       15.46   &       179.44   &            0.749   \\
				\hline
				         3 &      1.426 &      1.432 &      1.440 &       3.41 &      39.57 &              0.49 \\

				         4 &      1.423 &      1.426 &      1.431 &       2.10 &      24.44 &              0.36 \\

				         5 &      1.422 &      1.424 &      1.427 &       1.14 &      13.22 &              0.27 \\

				         6 &      1.421 &      1.423 &      1.424 &       0.69 &       7.98 &              0.23 \\
				\hline
		\end{tabular}
	\end{center}
	\caption{Optimized geometries of \tPA, PAc and $(n,n)$ CNTs with
$x=0.3$. 
}
	\label{tab:tubes_geom_etot}
\end{table}

The stability of these deformations was evaluated by calculating the total energy gain achieved by deforming the tube. The undistorted geometry was chosen such that it has equal A and B bond lengths, a common feature of aromatic systems. Note that within LDA and PBE, armchair carbon nanotubes also exhibit equal bond lengths. The structure was permitted to relax, under the above mention condition (A=B). This undistorted structure reaches its lowest total energy state when $A^{(0)} = B^{(0)} = \textrm{mean}(A^{(per)},B^{(per)})$. The bond-length C was allowed to relax but it was found to remain the same as in the fully-optimized structure. We calculated the total energy difference, $\Delta E_\textrm{tot}$, between the distorted and undistorted structures for all tubes. We report the results of this analysis in Table \ref{tab:tubes_geom_etot}. The stability of the deformations decreases rapidly with increasing tube diameter. Using these energy differences as an estimation of the temperature $T_{PD}$ at which the PD would occur, we find that it corresponds to a few kelvins for the nanotubes. Our estimates are between what was predicted  by Bohnen \emph{et al.} \cite{Bohnen:2004} and Conn\'etable \emph{et al.} \cite{Connetable:2005}, i.e. a near room temperature transition for the (3,3) nanotube, and what was suggested by Mintmire \emph{et al.} \cite{Mintmire:1992} and Piscanec \emph{et al.} \cite{Piscanec:2007}, i.e. $T_{PD}< 1$ K.

To our knowledge, the Peierls distortion in carbon nanotubes has never been directly and conclusively observed in experimental data, owing much from the fact that most readily available nanotubes are larger then the one studied here. As we have shown, the stability of such distortions decreases rapidly with increasing tube diameter. As such, it is comprehensible that small perturbations are enough to wash out this effect. Nonetheless, our results seems to concur with the resistance measurements of Shea \emph{et al.} \cite{Martel:2000}, which found a sharp increase in resistance at $T = 0.8K$ for a bundle of tubes. The diameter of the most abundant nanotube in the sample was $1.4$ nm, corresponding to a (10,10) tube. Extrapolating our results to such a tube, using a simple exponential fit \cite{note2}, we find $T_{PD} = 0.9$ K.

{\em Connection with the response function:} Within linear response theory, the Peierls instability can be understood as a pole in the response function $\chi(r,r')$ or equivalently as a zero eigenvalue in the $\chi^{-1}(r,r')$ spectrum. 
In a DFT approach, the real response of the system is related to the response function of the non-interacting Kohn-Sham system denoted by $\chi^s(r,r')$ by the following relation:
\begin{equation}
	 \frac{1}{\chi(r,r')}=\frac{1}{\chi^s(r,r')}-\left(\frac{1}{|r-r'|}+f_{xc}(r,r')\right),
\end{equation}
where $f_{xc}(r,r')$ is the exchange-correlation kernel. For a one-dimensional system, the Kohn-Sham non-interacting electron system exhibit a Peierl instability at a wave vector $q=2k_F$. The Coulomb part of the previous equation has a negative contribution. Since the response function $\chi^s(r,r')$ is negative definite, this Coulomb part will make the spectrum of $\chi(r,r')$ more negative. Hence it tends to decrease any instabilities in Kohn-Sham non-interacting electron system. On the other hand, the exchange-correlation kernel, since it is always negative, should attenuate/cancel the Coulomb part and preserve the Peierls instability. Therefore, the exchange-correlation kernel should contain a long-range contribution of the form $-\frac{\alpha}{q^2}$, where $\alpha$ is a material dependent parameter. We note that a kernel of the form $\chi(q)=-\frac{\alpha}{q^2}$ has been suggested for TDDFT calculations in order to obtain correct optical spectra \cite{Reining:2007}. Here we find that such response functions are important for the ground-state geometry in the system studied in this paper.

Based on this discussion, it is not surprising that exact exchange helps to reestablish the Peierls instability. Essentially, the direct Coulomb interaction alone would diminish the Peierls instability since it would tend to refrain electrons to localize in a same region. The local and near local functionals (LDA, PBE and HSE) tend to a constant has $q\rightarrow 0$. Hence, they fail to produce an appreciable Peierls distortion as they cannot cancel the Coulomb contribution. The inclusion of some exact-exchange restores a correct long-wave dependence of the kernel. The amount of exact-exchange must be chosen as to give the correct $\alpha$ parameter, which unfortunately will depend on the system studied. In this paper, we suggest that 30\% of exact-exchange is nominal for the study of 1D conjugated carbon systems such as \emph{t}-PA, PAc and armchair nanotubes.

{\em Conclusion:} Our DFT calculations show that hybrid functionals are better suited for the description of the properties of \emph{trans}-polyacetylene than local- or semi-local functionals (LDA or GGAs), and hence could be more appropriate to describe the geometry of carbon nanotubes. Using a hybrid functional derived from PBE, we showed that the distortion amplitude and the band gap increases with the amount of exact exchange included in the functional. Using the value of $ x $ that best describes the \tPA\ geometry, we found that small radius \emph{armchair} nanotubes should exhibit small lattice distortions that decrease with increasing tube diameter. Our particular choice of functional appears appropriate since it recovers the graphite bond lengths for larger tubes, in addition to describing the bond length alternation in \tPA\ correctly. We estimated the Peierls transition temperature to be of the order of few degrees K. 

We would like to thank Richard Martel helpful discussions and his assistance and Jeffrey Frey for providing the TubeGen Utility. Financial support through FQRNT and NSERC as well as computational resources made available by the R\'eseau qu\'eb\'ecois de
calcul de haute performance (RQCHP) are gratefully acknowledged.

\bibliography{biblio}

\begin{thebibliography}{10}
\bibitem{note2}%
  \BibitemOpen
  \bibinfo {note} {An exponential fit was used because it corresponds to the
  expected form of the Peierls transition temperature, as seen in Eq. 6.24 of
  reference \cite{sadovskii}.}%
  \bibAnnoteFile{Stop}{note2}%
\end{thebibliography}

\end{document}